\documentstyle[12pt,epsf,epsfig]{article}
\def\beq{\begin{equation}}
\def\eeq{\end{equation}}
\def\bea{\begin{eqnarray}}
\def\eea{\end{eqnarray}}
\def\bq{\begin{quote}}
\def\eq{\end{quote}}

\def\gappeq{\mathrel{\rlap {\raise.5ex\hbox{$>$}}
{\lower.5ex\hbox{$\sim$}}}}

\def\lappeq{\mathrel{\rlap{\raise.5ex\hbox{$<$}}
{\lower.5ex\hbox{$\sim$}}}}

\def\Toprel#1\over#2{\mathrel{\mathop{#2}\limits^{#1}}}

\begin{document}
\pagestyle{empty}
\begin{flushright}
{CERN-TH/2001-023}\\
hep-ex/0112004\\
\end{flushright}
\vspace*{5mm}
\begin{center}
{\bf PHYSICS MOTIVATIONS FOR FUTURE CERN ACCELERATORS}~$^*$ \\
\vspace*{1cm}
{\bf Albert De Roeck,  John Ellis  and  Fabiola Gianotti}\\
\vspace{0.3cm}
CERN, CH - 1211 Geneva 23 \\
\vspace*{2cm}
{\bf ABSTRACT} \\ \end{center}
\vspace*{5mm}
\noindent
We summarize the physics motivations for future accelerators at CERN.  We
argue that (a) a luminosity upgrade for the LHC could provide good physics
return for a relatively modest capital investment, (b) CLIC would provide
excellent long-term perspectives within many speculative scenarios for
physics beyond the Standard Model, (c) a Very Large Hadron Collider could
provide the first opportunity to explore the energy range up to about 30
TeV, (d) a neutrino factory based on a muon storage ring would provide an
exciting and complementary scientific programme and a muon collider could
be an interesting later option.

\vspace*{2.5cm}
$^*$ {\it Prepared for the CERN Scientific Policy Committee in September
2001, and presented to the CERN Council in December 2001}
\vspace*{2.5cm}

\begin{flushleft} CERN-TH/2001-023\\
December 2001
\end{flushleft}
\vfill\eject

\setcounter{page}{1}
\pagestyle{plain}

\section{Introduction}

The central motivation for any CERN accelerator beyond the LHC must be
provided by physics beyond the present Standard Model. LEP has tested the
Standard Model very precisely, but has provided no direct evidence for
physics beyond it. Atmospheric and solar neutrino oscillation experiments
have provided the first experimental evidence for physics beyond the
Standard Model, and long-baseline experiments using accelerator neutrino
beams are underway to verify and explore this new physics. Recently,
the Brookhaven experiment on the anomalous magnetic moment of the muon has
reported a hint of a possible discrepancy with the Standard Model.
Nevertheless, the possible direction of experimental physics beyond the
Standard Model remains uncertain, and any prediction which experiments
will be interesting for CERN after the LHC and the CERN-Gran Sasso
neutrino project must be largely speculative.

The most prominent areas where the Standard Model leaves unsolved problems
include the origins of the particle masses, the variety of particle types
(flavours), the unification of the particle interactions, and a consistent
quantum theory of gravity.  All of these problems should ultimately be
resolved in a `Theory of Everything', and the long-term goal of particle
physics is to discover and establish such a theory.  Resolutions of the
more immediate problems will provide some of its ingredients.

Particle masses are believed to be due to a Higgs boson, and LEP provided
last year a hint that it might weigh about 115 GeV.  The masses generated
by the Higgs mechanism are, however, destabilized by quantum effects,
unless new physics is invoked.  Favoured theoretical scenarios for
stabilizing particle masses have been the idea that the Higgs boson might
be composite, or that it might be accompanied by supersymmetric particles,
which have identical quantum numbers to the known particles, but have
spins differing by half units.  More recently, inspired by string theory,
it has been suggested that there might be additional spatial dimensions
beyond the three we know, and that these might remove the need for a large
hierarchy between the intrinsic scale of gravity and the observed particle
masses~\cite{aaa}.

These scenarios all predict new physics at the TeV energy scale, which
will initially be explored by the LHC.  If confirmed, the Brookhaven
measurement of a possible discrepancy in the anomalous magnetic moment of
the muon would also hint at new physics at the TeV energy scale, and
interpretations based on supersymmetry and lepton compositeness have been
proposed.  On the other hand, there is no clear hint of the energy scale
where new flavour physics might be revealed, and the natural scales of
grand unification and quantum gravity are beyond direct reach at
foreseeable colliders, unless there are large extra dimensions.  However,
these might be tested indirectly, e.g., by the relations they predict
between low-energy parameters, or by novel phenomena such as neutrino
oscillations.

Despite the lack of direct accelerator evidence for physics beyond the
Standard Model, LEP has provided some indirect indications which future
directions might prove fruitful, and disfavoured some others.  For
instance, the precision electroweak data suggest the existence of a
relatively light Higgs boson, favouring some weakly-coupled model of
electroweak symmetry breaking, such as supersymmetry or a theory with
large extra spatial dimensions.  On the other hand, these data disfavour
strongly-interacting models with composite Higgs bosons, though we do not
ignore such a possibility in the following. If there is indeed a light
Higgs boson with mass close to the experimental lower limit set by LEP,
the Standard Model cannot survive unchanged until the Planck or
unification scale, and, more specifically, new physics would be {\it
required} below about $10^5$ to
$10^6$~GeV if the LEP hint of a 115 GeV Higgs boson were to be
confirmed~\cite{bb}.

Even if the existence of supersymmetry or large extra dimensions were to
be discovered by the LHC, full understanding of the theory behind would
require additional measurements with a complementary collider.  For
example, what breaks supersymmetry, or what fixes the sizes of the extra
dimensions?  Plans for future CERN accelerators beyond the LHC should bear
such questions in mind.

\begin{table}
\begin{center}
\begin{tabular}{|l |l |l|l|l|l |l|}
\hline
Process & LHC & LC & SLHC & VLHC &\multicolumn{2}{c|}{CLIC} \\
& 14 TeV & 0.8 TeV & 14 TeV & 200 TeV& 3 TeV & 5 TeV \\
\hline
squarks & 2.5 & 0.4& 3 & 15 & 1.5 & 2.5 \\
sleptons & 0.34 & 0.4 &  & & 1.5 & 2.5\\
$Z'$ & 5.4 & 8 &6.5 & 30 & 20 & 30\\
$q^*$ & 6.5 & 0.8 & 7.5 & 70 & 3 & 5 \\
$l^*$ & 3.4 & 0.8 & & & 3 & 5 \\
Extra two dimensions & 9 & 5 -- 8.5 & 12 & 65 & 20 -- 33 & 30 -- 55 \\
$W_LW_L$ & 3.4$\sigma$ & & $>3.4\sigma$ & 30$\sigma$ & 70$\sigma$ & 
90$\sigma$ \\
TGC (95\%) & 0.0014 & 0.0004 & 0.0006 &0.0003 & 0.00013 & 0.00008 \\
$\Lambda$ compos. & 35 & 100 & 50 & $130$ & 300 & 400 \\
\hline
\end{tabular}
\end{center}
\caption{\it Comparison of physics reaches with different colliders.
The integrated luminosities assumed are 100 fb$^{-1}$ for the LHC and
VLHC, 500 fb$^{-1}$ for the 800 GeV LC, and 1000 fb$^{-1}$
for the SLHC and CLIC,
corresponding in each case to one full year of running at nominal
luminosity.  Most of the numbers quoted are in TeV, but for strong $W_L
W_L$ scattering the numbers of standard deviations are listed, and in 
Triple Gauge Coupling (TGC) case a pure number is given for
$\lambda_\gamma$. In some cases,
the sensitivities of hadron colliders to electroweak physics have not yet
been evaluated accurately, and the corresponding entries are left blank.
Many of the numbers given for the SLHC, VLHC and CLIC are still
provisional.}
\end{table}

 The principal options~\cite{cc} for possible CERN accelerators beyond the
LHC
that we have considered include (i) an upgraded LHC with $E_{cm} = 14$~TeV and
a luminosity of $10^{35}$~cm$^{-2}$s$^{-1}$ that we call the SLHC, (ii) a
higher-energy hadron-hadron collider with $E_{cm}$= 100 to 200~TeV (the
Very Large Hadron Collider - VLHC)~\cite{dd}, (iii) a linear $e^+e^-$ collider with $E_{cm}$ = 3 to 5~TeV,
such as the CLIC collider that is being studied actively at
CERN~\cite{ee}, and
(iv) a muon collider (MC) with $E_{cm} < 4$~TeV.  We couple the latter with
other  physics opportunities offered by muon storage rings, such as a
neutrino factory and Higgs factories~\cite{ff,gga}.

\section{Physics Reach of the LHC}

We consider in the following a variety of possible physics topics beyond
the Standard Model, including a range of benchmark scenarios for
supersymmetric models consistent with the present experimental limits from
LEP and elsewhere.  We list in the Table the estimated physics reach of
the LHC~\cite{hh,ii} for these various types of physics beyond the
Standard
Model, including the accessible ranges for the masses of the possible
supersymmetric partners for some known particles, of a new neutral weak
boson $Z'$ analogous to the well-known $Z$, of excited quark and lepton states
$q^*$ and $l^*$, the accessible size $R_D$ of a pair of extra dimensions
(which we parametrize by the energy scale $M_D$ equivalent to $1/R_D$),
the sensitivities
to strong interactions between pairs of $W$ bosons parametrized by a 1~TeV
Higgs boson and to deviations from the Standard Model value for the Triple
Gauge Coupling (TGC) $\lambda_\gamma$, and the reach for a compositeness
scale $L$ at which
elementary particles might reveal internal structure.

As is well known, the LHC has a large reach for the strongly-interacting
partners of quarks and gluons (the squarks and gluinos), and would be
capable of reconstructing many of their intricate cascade decay modes.
However, its reach for non-strongly-interacting supersymmetric particles
such as the supersymmetric partners of leptons (sleptons) is more limited.
The set of simplified benchmark supersymmetric models we have
established~\cite{jj}
have universal supersymmetric mass parameters that are chosen to be
compatible with the various Higgs boson and sparticle limits from
LEP~\cite{aaa,jj,kk}.  The universality we assume is just one of many specific
variants of supersymmetry, and one of the primary tasks of future
accelerators may be to probe its assumptions.  The first panel of the
Figure illustrates crudely the fractions of the rich sparticle spectra in
these benchmark models that we estimate to be accessible to the LHC
experiments.  We see that the LHC discovers some supersymmetric particles
in all these scenarios, but never discovers them all.  For this reason,
the LHC alone will not be able to pin down all the details of
supersymmetry breaking.

\begin{figure}
\epsfig{figure=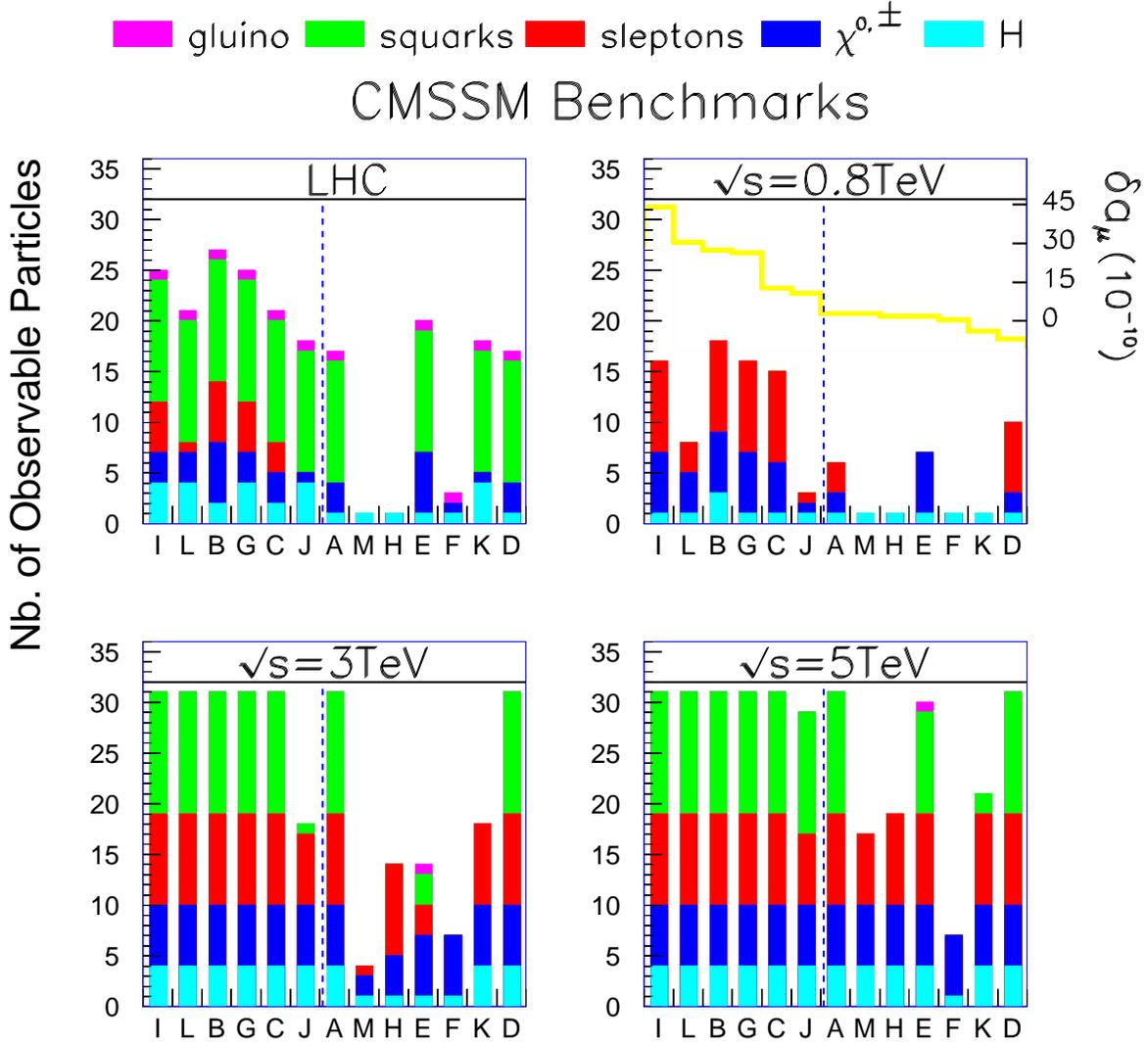,width=16cm}
\caption[]{\it Comparison of the capabilities of various colliders to 
observe
different species of supersymmetric particles, in a range of
supersymmetric models whose parameters are chosen to be compatible with
experimental constraints from LEP and elsewhere~\cite{jj,kk}.  The models
are
ordered by their degree of compatibility with the recent BNL measurement
of the muon anomalous magnetic moment, as indicated by the line in the top
right panel. We see that linear colliders complement the LHC via their
abilities to observe weakly-interacting sparticles, in particular. In most
of the restricted set of simplified models studied, CLIC (almost)
completes the supersymmetric spectroscopy initiated by the LHC.}
\end{figure}

Likewise, in models with large extra dimensions, the new physics may well
show up within the range shown in the Table, although mass scales up to
100~TeV are also possible. Also, if the LHC were to discover a Higgs boson
weighing about 115~GeV, but found no supersymmetric particles, the LHC
would not cover the energy range up to $10^5$ to $10^6$~GeV where new physics
should appear.

We infer from the Table and the Figure that, although the LHC has
excellent prospects of making fundamental discoveries beyond the Standard
Model, with mass reach well into the TeV range, it is unlikely to provide
complete answers to all the ensuing questions.

To complete the baseline for our subsequent discussion of future CERN
colliders, we also include in the Table a column showing the physics reach
of a linear $e^+e^-$ collider (LC) with $E_{cm} = 800$~GeV, such as TESLA,
the NLC or JLC.  Such a LC would complement the LHC in several respects, e.g., in
the search for new weakly-interacting physics.  This point is made in the
second panel of the Figure, where we see that such a LC would (in several
supersymmetric scenarios) observe non-strongly-interacting supersymmetric
particles that would not be visible at the LHC.  Moreover, a LC would
provide many detailed measurements of sparticle properties that would lie
beyond the scope of a hadron-hadron collider.

\section{Physics Opportunities with Possible Future Accelerators}

In this section we discuss the physics accessible to the possible
accelerator options introduced above, namely the SLHC, VLHC, CLIC and
muon storage rings.

\subsection{SLHC}

Preliminary estimates by the LHC project team~\cite{Evans}
indicate that it should be possible to increase the peak
luminosity from the design value of $10^{34}$~cm$^{-2}$s$^{-1}$ up to 
about
$10^{35}$~cm$^{-2}$s$^{-1}$  by  increasing the bunch intensity up to the
beam-beam limit,
replacing  the inner quadrupole triplets with larger aperture magnets to
reduce the $\beta$-value at the interaction points, and by halving the
bunch
spacing to 12.5 ns in order to preserve the luminosity lifetime.

On the other hand, more than a modest LHC energy upgrade  looks much more
difficult.  The present dipoles may ultimately sustain a 9 Tesla
field,
which would correspond to a maximum $E_{cm}$ = 15.2~TeV. Even if advances
in
magnet technology made a substantially larger field possible,  other
problems such as the extraction of the increased synchrotron radiation
power would present difficulties for fitting a significantly higher energy
machine into the LHC tunnel.

Although the ultimate LHC centre-of-mass energy with the present magnets
could be up to 15.2~TeV, the SLHC studies~\cite{lla,mm} presented
here~\cite{nn} have
conservatively assumed $L=10^{35}$~cm$^{-2}$s$^{-1}$ and  $E_{cm}$=14~TeV.
Running the LHC
detectors at  ten times the nominal LHC luminosity would cause
non-negligible radiation and occupancy problems, but solutions appear
feasible~\cite{oo}.  Here we only mention that one of the main
consequences of
the luminosity upgrade would be that a large part of the inner detectors
of both experiments, e.g., the innermost layers of both trackers and the
ATLAS Transition Radiation Tracker, would need to be replaced, in order to
maintain the required performance for bottom-quark tagging,
electron and tau measurements. On the other hand, calorimeters and the
external muon spectrometers should be less affected by the luminosity
increase, and should remain fully functional  with upgrades of smaller
scope.

The estimated physics reaches of the  SLHC for supersymmetry, $Z', q^*, l^*,
M_D$, strong WW scattering, TGC and $\Lambda$ are shown in the third
column of the Table.  In the discovery channels, we see that the 
SLHC reaches somewhat
further than the LHC, typically by about 20\%.  The order-of-magnitude
increase in statistics should also make possible significant improvements
in precision measurements, as indicated by the increased sensitivity to
the TGC shown in the Table.  Although these improvements are not dramatic,
this relatively modest upgrade of the LHC might be very interesting if the
LHC finds hints of some phenomena that need to be pinned down.  For
example, in some of our benchmark supersymmetric models, we expect gluinos
to weigh between 2.5 and 3 TeV, which is most likely in the range covered
by the SLHC but not the LHC.  Alternatively, in some supersymmetric models
some flavours of squarks might be heavier than others, and the SLHC could
complete the squark spectrum revealed by the LHC.

The results shown in the Table were obtained for  final states containing
objects  with very large transverse energy, such as jets, leptons, photons
or missing transverse energy.  Since the pile-up noise in the calorimeter
increases only by a factor of about 3 for a factor of 10 increase in
luminosity, and gives a negligible contribution to the measurements of
particles in the TeV range, calorimetric measurements of jets, electrons,
photons and
missing transverse energy at the SLHC should  not suffer from the higher
luminosity.  However, electron identification requires a combination of
calorimeter and tracker information, and the latter will only be available
if the LHC inner detectors are largely replaced.  Therefore, in order to
be conservative, and not knowing a priori what level of detector upgrade
will be technically and financially possible, for the studies presented
here we have ignored electrons and  considered only final states
containing muons, jets, photons and missing transverse energy.  It is
clear, however, that full detector functionality, including the
possibility of detecting and identifying all three lepton species, would
be essential to profit fully from the increased luminosity of the SLHC,
and to obtain more convincing results in the event of a discovery.

        The performance of a hadron collider with $E_{cm}$ = 28~TeV and $L
=
10^{34}$~cm$^{-2}$s$^{-1}$ has also been looked at, in order to estimate 
the available
physics as a function of energy as well as luminosity.  Such a machine
would extend the LHC physics reach by a factor between 1.5 and 2,
depending on the channel studied.  For example, a signal from
strongly-interacting W pairs should be observable at the
five-standard-deviation level with only 50~fb$^{-1}$ of integrated luminosity.
However, it is not known currently how such a machine could be housed in
the LHC tunnel, since it would require very high-field magnets and new
ways to deal with synchrotron radiation.

\subsection{VLHC}

Studies have been made in the United States of the physics offered by a
100-200~TeV 
hadron collider, the VLHC~\cite{dd}.   Some estimates of the reach
of
this machine for $L = 10^{34}$~cm$^{-2}$s$^{-1}$ are also presented in the 
Table,
providing useful information about the physics interest of the energy
regime beyond the LHC/SLHC.

Primarily for geographical reasons, we do not consider a VLHC to be the
most likely option for CERN's future~\cite{cc}.  The most recent design
study
for a VLHC 
presents a machine with a circumference of 233 Km for 
$E_{cm}$ = 200~TeV.  The maximum size of
ring that could be accommodated in the Geneva basin is guessed to have a
circumference of about 80 Km, and the closest that a larger machine could
be placed is probably beyond the Jura mountains in France.  VLHC technical
studies have been underway for several years in the United States.  The
main engineering issue is to find technically and economically viable
solutions for the machine components, such as the magnets and cryogenics,
and the tunnel.

Just as the LHC will be the first machine to enter and explore the TeV
energy range, the VLHC will be able to explore the energy range up to
about 30-40~TeV in the centre of mass for constituent collisions.  In the
absence of input from the LHC, it is not yet possible to make a compelling
case for any particular scenario for new physics at the 10~TeV scale, in
contrast to the TeV energy scale.  However, we can envisage several
possible scenarios emerging from the LHC data which would justify strongly
the need for  such a machine~\cite{nn}.  A few examples are:
\begin{itemize}
\item
The LHC discovers supersymmetry, and supersymmetry is mediated by new
gauge interactions.  In this case,  by measuring several sparticle masses
and the lifetime of the next-to-lightest supersymmetric particle, the LHC
experiments should be able to constrain, to within 30\% or so, the scale at
which supersymmetry breaking is communicated to the visible world.  If
this scale is in the energy range up to a few tens of TeV,  then a VLHC
could produce directly the corresponding new particles.

\item
The LHC discovers supersymmetry, and observes the squarks of the third
generation, the supersymmetric partners of the top and bottom quarks, but
not the squarks of the first two generations. This is possible in
so-called inverted-hierarchy models, where the squarks of the first two
generations can weigh several TeV without creating  naturalness problems.
We will then know that such squarks should exist, and a VLHC, which could
produce squarks weighing up to about 15 TeV,  is the only presently
foreseen machine able to observe them.
\item
The LHC finds evidence for quark compositeness by observing a
significant excess of centrally produced high-$E_T$ di-jet events above
the
Standard Model expectation.  It can be seen from the Table that the
compositeness scale $\Lambda$ would then be in the energy range up to a
few tens
of TeV. Then a VLHC could probe the production of new particles, such as
excited quarks, giving more direct and conclusive evidence for
compositeness.
\item
The LHC finds a hint for strong WW scattering, in which case a VLHC
could study it with much higher statistics.
\item
The LHC finds evidence for large extra spatial dimensions, for instance
by observing a significant signal in final states with jets and missing
transverse energy.  It can be seen from the Table that the fundamental
scale of gravity would then be in the region of a few tens of TeV, and  a
VLHC should again be able to probe directly the scale of new physics.
\end{itemize}

\subsection{CLIC}

As already mentioned, linear electron-positron collider (LC) projects are
being pursued actively in Germany~\cite{pp}, the U.S.~\cite{qq} and
Japan~\cite{rr}, as well as at CERN~\cite{ee}.  The first-generation LC
projects being pursued elsewhere typically aim at centre-of-mass energies
in the range 0.5 to 1~TeV, and we have chosen $E_{cm}$ = 800~GeV as a
standard~\cite{pp} for comparison with the LHC and CERN's future options.

The physics reach for such a LC is inferior to the LHC for new
strongly-interacting particles.  On the other hand, as also seen in the
Figure, such a first-generation LC would be largely complementary to the
LHC, offering, for example, excellent opportunities to study in detail any
light Higgs boson and make precise measurements of any supersymmetric
particles within its reach, in particular the supersymmetric partners of
weakly-interacting particles~\cite{pp}.  However, there is no guarantee
that
any supersymmetric particles would be accessible to a LC with 
$E_{cm}$ = 800~GeV, and such a LC would not in general be able to complete the
supersymmetric spectroscopy, as seen in the second panel of the Figure.
We assume that some such sub-TeV LC will be built somewhere in the
world, and that CERN should envisage constructing a higher-energy
lepton-lepton collider, such as CLIC or a muon collider.

CERN and its partner laboratories have been developing for several years
the CLIC double-beam technology for generating high accelerating gradients
in the range 100 to 200~MV/m, enabling a higher-energy LC to be
constructed in a tunnel of length similar to that proposed for a
lower-energy LC with a lower accelerating gradient~\cite{ee}. Preliminary
geological studies indicate that a tunnel up to about 35 Km long could be
accommodated in good rock between the Jura mountains and Lake Geneva, in
the immediate neighbourhood of the present CERN site~\cite{cc}. This would
be
sufficient to accommodate a CLIC machine with centre-of-mass energy up to
3 or 5~TeV.  CLIC would be able to provide polarized beams, which are
useful for several physics topics such as sparticle studies and searches
for extra dimensions (see the second numbers in the last two columns of
the sixth row of the Table), and would also have options for $e \gamma$
and $\gamma \gamma$ collisions, which have better reaches for some physics
topics.

We see in the Table that CLIC with $E_{cm}$ = 3 to 5~TeV has an impressive
physics reach for all the physics topics studied~\cite{clicphysics,jj},
with many
capabilities beyond those of the LHC and a first-generation LC.  Moreover,
we see in the Figure that CLIC would be able to complete much of the
sparticle spectroscopy opened up by the LHC, going significantly beyond
the reach of a lower-energy LC.  As an example, CLIC may be the first
machine able to disentangle the different squark flavours and complete the
spectra in most of our simplified benchmark models if it attains $E_{cm}$ = 5~TeV.  We think it is important not to lose sight of this ultimate goal for
CLIC.  CLIC also has excellent potential for many of the other physics
topics in the Table, such as extra dimensions, strong WW scattering,
measuring the TGC, and lepton compositeness.  For example, CLIC would
enable multi-TeV Kaluza-Klein resonances in extra dimension scenarios to
be studied in great detail.

The experimental environment at CLIC offers several challenges, as
compared to  a lower-energy LC.  For example, the colliding beams are
expected to radiate energetic photons much more strongly, reducing and
smearing the nominal centre-of-mass energy, and leading to an imbalance in
the visible momentum in the final state. Also, photon-photon collisions
are expected to be relatively more copious than at a lower-energy LC, and
large incoherent creation of electron-positron pairs is also expected.
Dealing with these problems requires close collaboration on the
machine-detector interface, which has already started within the CLIC
physics study group.

Initial analyses do not reveal any showstopping problems created by the
relatively large beam energy spread inherent to CLIC at high energy and
luminosity.  For example,  direct studies of a heavy $Z'$ boson analogous to
the known $Z$ would be very easy at CLIC, enabling many of its properties to
be measured with per mille precision~\cite{ss}.  Studies have also been
made of
the production of pairs of supersymmetric particles~\cite{tt}. Their
missing-energy signatures would be quite distinctive, and the thresholds
for their production could  also be measured easily.  In addition to
discovering and measuring precisely new supersymmetric particles, CLIC
would also be able to distinguish and measure more precisely sparticles
discovered at the LHC.

There are several ways in which the extra information provided by CLIC
could prove to be essential.  A few examples are:
\begin{itemize}
\item
The LHC discovers supersymmetry and measures the masses of a few
supersymmetric particles, with the first-generation LC perhaps finding and
measuring some more.  As previously mentioned, in the simplified benchmark
supersymmetric models we have studied, CLIC would complete the
supersymmetric spectrum and enable detailed measurements to be made. Thus
it might cast light on the mechanism of supersymmetry breaking, which
might be a window into string physics.
\item
The LHC finds evidence for large extra dimensions. In this case, the
combination of data at different CLIC energies will enable the number and
size of the extra dimensions to be determined independently.
· The LHC finds evidence for quark compositeness.  In this case, CLIC
could reveal lepton compositeness, completing the revelation of a new
layer of fundamental structure.  It might well also be that excited
leptons would appear at CLIC even if excited quarks have not appeared at
the LHC, if CLIC attains 5~TeV.
\item
The Higgs boson weighs 115~GeV, as suggested by LEP, but the LHC finds
no supersymmetry or evidence for quark compositeness.  CLIC would be able
to measure the Higgs self-coupling with better than 10\% precision,
enabling the
effective potential to be mapped, would observe any charged Higgs bosons
weighing up to 2~TeV or so, and would provide new opportunities to observe
the heavier neutral Higgs bosons in supersymmetric models.  Moreover, we
see from the Table that CLIC is best placed to find compositeness over
essentially all the energy range up to about $10^5$ to $10^6$~GeV, where new
physics must appear if the Higgs boson weighs 115~GeV.
\item
The LHC finds a hint of strong WW scattering, in which case CLIC could
study it with high statistics and precision~\cite{uu}.
\end{itemize}

\subsection{Muon Storage Rings and Colliders}

        Another possible option for CERN's future is to develop a complex
of muon storage rings.  The first component would be a high-intensity
proton driver, for example a superconducting proton linac (SPL) that would
reuse LEP radiofrequency cavities~\cite{vv}.  This could also make
possible
interesting upgrades for other CERN facilities, ranging from ISOLDE to the
CERN-Gran Sasso neutrino beam and the LHC. The SPL would also offer
interesting possibilities in short-baseline neutrino physics and studies
with stopped muons.

The next steps in a programme of physics with muons would be to capture
them, cool them, and accelerate them in recirculating linacs.  The first
physics option offered would be simply to store the energetic muons,
allowing them to decay producing intense and well-understood beams of
electron and muon neutrinos -- a neutrino factory~\cite{ff,gga,ww}.  This
would
offer interesting physics opportunities in the search for long-baseline
oscillations between electron and muon neutrinos and matter effects on
neutrino oscillations.  The most exciting opportunity may be the search
for CP violation~\cite{ww}. A neutrino factory would be the logical next
step
after the current generation of long-baseline neutrino experiments, and
might provide a unique window on grand unification via the lepton sector.

        If the technique of muon cooling could be refined, the next
options might be to collide muons in one or more Higgs factories,
exploiting the direct-channel $\mu^+\mu^- \rightarrow H$ production mechanism.  Such a MC
would be a unique machine for measuring the Higgs line shape, as LEP did
for the $Z$.  It would also offer ideal opportunities to study CP violation
in the Higgs sector, as expected in some supersymmetric models.  In the
longer run, one could also envisage building  a high-energy MC, which
would probably be limited to $E_{cm} < 4$~TeV by the neutrino radiation
hazard.
Its physics reach would be equivalent to that of CLIC for the same
$E_{cm}$ and
luminosity, and the smaller beam-energy spread and better energy
calibration would confer some advantages on a MC, though the absence of $\mu
\gamma$ and $\gamma \gamma$   collisions might be a handicap.  There are some physics
scenarios in which colliding muons may be preferable to colliding
electrons, for example in Higgs physics or in certain supersymmetric
models that violate R parity.  On the other hand, muon decays will provide
important challenges to detector builders, and large controlled beam
polarization will not be available, unlike at CLIC.

Since there are many technical hurdles on the route to a MC, which surely
has a longer development time-scale than CLIC, we do not discuss it
further, except to comment that a muon storage ring complex, including a
high-energy MC, could fit comfortably within the area of the existing CERN
accelerators.

\section{Summary}

We have discussed briefly possible future high-energy accelerator options
for CERN, and note the following preliminary impressions.

\begin{itemize}
\item{(a)}  The SLHC presents interesting possibilities that may provide 
good
physics return for the relatively modest capital investment required to
upgrade the LHC luminosity and the LHC detectors.

\item{(b)}  CLIC provides excellent long-term perspectives within all the
speculative scenarios for physics beyond the Standard Model that we have
considered, and is therefore a very attractive option for CERN's long-term
future beyond the LHC.  CLIC would complement and go beyond the LHC, being
an ideal machine to study heavy new weakly-interacting particles, and
could fit comfortably in the neighbourhood of the CERN site.
\item{c)}  A neutrino factory based on a muon storage ring would provide an
exciting and complementary scientific programme, probing grand unified
theories, but a high-energy muon collider is not a prospect for the near
future.
\item{d)}  A decision on the construction of a VLHC, which is unlikely to fit in
the neighbourhood of the CERN site,  could only be taken after  a few
years of  LHC running at the design luminosity.  Should the LHC data give
indications that the next energy scale, e.g., that of supersymmetry
breaking, compositeness or extra dimensions, lies around 50 TeV  or below,
then  a VLHC  would be the ideal  machine to produce directly the
corresponding new particles and probe their interactions.

\end{itemize}

In conclusion, we comment that many open questions will be answered by the
LHC, and possibly other experiments, before the decision about CERN's
following high-energy accelerator needs to be finalized.  The LHC will
certainly provide many clues to the `Theory of Everything', but further
accelerator experiments will be needed to identify it.  Quite possibly,
some completely new phenomenon will be discovered at the LHC that will be
more exciting than any of the theoretical speculations discussed here.
This brief survey has indicated that there are already many strong options
for new CERN accelerators beyond the LHC, suitable for addressing the
questions it will leave open.

CERN has, in the LHC, a forefront project that will establish its
leadership in high-energy physics. The organization should have similar
ambition for its projects after the LHC.

\section*{Acknowledgements}

We thank M. Battaglia, U. Baur and A. Blondel for their help.

\end{document}